\documentclass[aps,prb,showpacs,twocolumn,superscriptaddress]{revtex4-1}

\usepackage{graphicx}
\usepackage{amsmath,amssymb}
\usepackage{hyperref}
\usepackage{natbib}
\usepackage{bm}

\begin{document}

\title{Time evolution of Floquet states in graphene}

\date{\today}
\author{F.~Manghi}
\affiliation{Dipartimento di Scienze Fisiche, Informatiche e Matematiche, Universit\`a di Modena e Reggio Emilia, Via Campi 213/A, I-41125 Modena, Italy }
\affiliation{CNR - Institute of NanoSciences - S3}
\author{M.~Puviani}
\affiliation{Dipartimento di Scienze Fisiche, Informatiche e Matematiche, Universit\`a di Modena e Reggio Emilia, Via Campi 213/A, I-41125 Modena, Italy }
\author{F.~Lenzini}
\affiliation{Dipartimento di Scienze Fisiche, Informatiche e Matematiche, Universit\`a di Modena e Reggio Emilia, Via Campi 213/A, I-41125 Modena, Italy }

\begin{abstract}

Based on a solution of the Floquet Hamiltonian we have studied the time-evolution of electronic states in graphene nanoribbons   driven out of equilibrium by  time-dependent electromagnetic  fields   in different regimes of intensity, polarization and frequency.  We show that the time-dependent  band structure contains many unconventional features   that are not captured by considering the Floquet eigenvalues alone. By analyzing the evolution in time of the state population we have   identified regimes for the emergence of  time-dependent edge states responsible of  charge oscillations across the ribbon.

\end{abstract}


\maketitle

If a time periodic field is applied to electrons in a periodic lattice the Bloch theorem can be applied twice, both in space and in time. This is the essence of the Floquet theory \cite{Shirley1965,Faisal1997,Grifoni1998} that has recently attracted a  renewed interest for its ability to describe topological phases in driven quantum systems.\cite{Goldman2016,Sentef2015,PhysRevA.92.023624,Cayssol2013b}
The  discovery that circularly polarized light may induce non trivial topological behaviour in materials that would be standard in static conditions \cite{Usaj2014,PhysRevX.3.031005,Lindner2011,Oka2009} has  opened the way to the realization of  the so-called Floquet topological insulators, where a topological phase may be engineered and manipulated  by  tunable controls such as  polarization, periodicity and amplitude of the external perturbation.

In the presence of a continuous periodic driving electrons are in a non-equilibrium steady state characterized  by a periodic time dependence of the wavefunctions and, consequently, of  the expectation values of any observable.
In this paper we focus on this time dependence, looking for the time evolution of some relevant observables such as energy and charge density.  We will consider the prototypical case of graphene that under the influence of circularly polarized light exhibits in its Floquet band structure  the distinctive characteristics of a topological insulator, namely a gap in 2D and linear dispersive edge states in 1D.\cite{Usaj2014,PhysRevB.89.121401,PhysRevLett.107.216601,Oka2009} How these characteristics affect the time behaviour of some observables will be our focus.

Under a periodic driving the non-equilibrium steady states, solutions of the time-dependent Schroedinger equation
\begin{equation*}\label{tds}\left(\hat{H}(\textbf{r},t)  -\imath \frac{\partial}{\partial t}\right)\psi(\textbf{r},t)= 0 ,\end{equation*}
  evolve in time as
\begin{equation}\label{psit}
\psi(\textbf{r},t)= e^{-\imath  \epsilon_{\alpha}t} \phi(\textbf{r},t)
\end{equation}
where $ \phi(\textbf{r},t)$ is periodic in time and $\epsilon_{\alpha}$ - the Floquet  quasi-energies - are the eigenvalues of an effective    Hamiltonian $\hat{H}^F  \equiv \hat{H}  -\imath \frac{\partial}{\partial t}$ - the so-called Floquet Hamiltonian:
\begin{equation}
\label{floqueteigen}
\hat{H}^F \phi_{\alpha}(\textbf{r},t) = \epsilon_{\alpha} \phi_{\alpha}(\textbf{r},t)  .
\end{equation}
Here $\hat{H}(\textbf{r},t) $ is the full Hamiltonian of the driven system
\begin{equation}\hat{H}(\textbf{r},t) = \hat{H}_0(\textbf{r})+ \hat{V}(\textbf{r},t) \end{equation} with $H_0(\textbf{r})$ the static hamiltonian
and $V(\textbf{r},t)$ the external periodic driving.
\begin{figure}[h]
\centering \includegraphics[width=5cm]{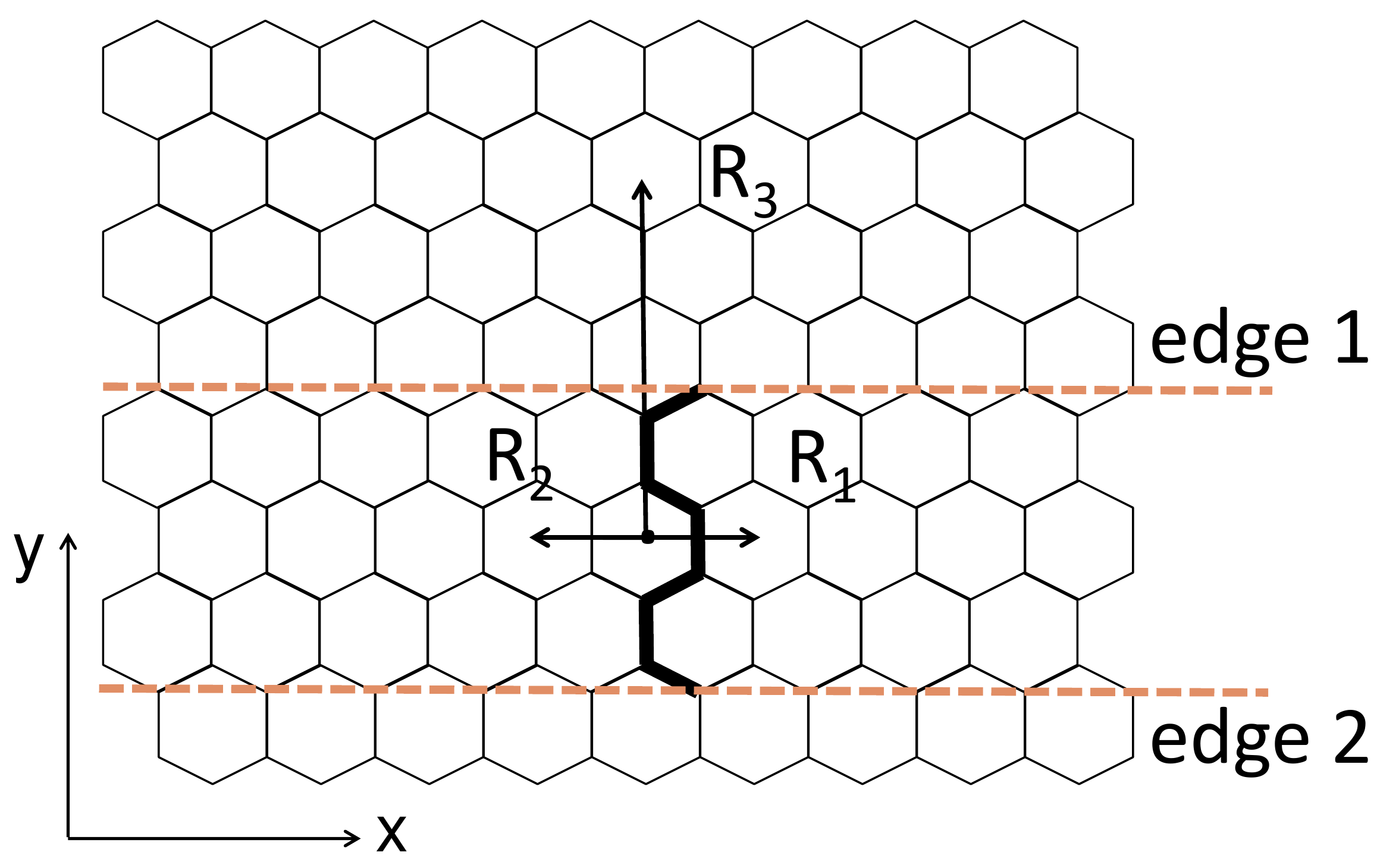}
\caption{\label{geo} Geometry of a zig-zag honeycomb ribbon. Heavy lines indicate the  unit cell for a 8-atom wide ribbon. $R_1, R_2$ are the lattice vectors associated to 1D translation symmetry. $R_3$ is the  lattice vector to be  used to reproduce the 2D lattice starting from the present unit cell (see text).}
\end{figure}
The factorization in eq. \ref{psit}  is exact and represents the temporal analogue of the Bloch theorem.
Being $\phi_{\alpha}(\textbf{r},t)$ time-periodic it can be expressed as a Fourier series
\begin{equation}\label{FT}
  \phi_{\alpha}(\textbf{r},t)=\sum _{n=-\infty}^{\infty} B_{\alpha n}(\textbf{r}) e^{-\imath n \Omega t} .
\end{equation}
where in turn $B_{\alpha n}(r)$ can be expanded on a complete set, for instance  on  a localized basis
\begin{equation}\label{basis}
    B_{\alpha n}(\textbf{r})= \sum_{i}^{N} C_{n i}^{\alpha} \chi_i(\textbf{r})
\end{equation}
with $i$ a site index, $N$ the number of sites in the unit cell and $\chi_i(\textbf{r})$ the localized orbitals \footnote{Index $\alpha$ includes k-vector, band index and Floquet mode.}.
In practice the Fourier expansion  is truncated to include a finite number of modes, up to a sufficiently large $n_{max}$ whose value depends obviously on $\Omega$. This allows to formulate the eigenvalue  problem in eq. \ref{floqueteigen} in a standard matrix form whose eigenvalues  turn out to be replicas of the static band structure with gaps opening  at their crossing points.\cite{Faisal1997,Usaj2014}

The field-free hamiltonian of graphene is described in the  tight-binding scheme  with  a single hopping parameter $J \simeq 2.8 eV$ between nearest neighbor sites, reproducing the well known Dirac-like valence and conduction bands \cite{TBgraphene}. In the presence of the oscillating field  described by the vector potential $\textbf{A}(t)$, the hopping
between neighboring sites is modified according to  Peierls' substitution \cite{PhysRevLett.108.225303,Hofstadter1976}
\begin{equation}\label{hop}
   \tilde{J}_{i j}(t)=J e^{i\textbf{A}(t)\cdot (\textbf{r}_{j}-\textbf{r}_i)}
\end{equation}
\onecolumngrid
\begin{center}
\begin{figure}[h]
\centering \includegraphics[width=14cm]{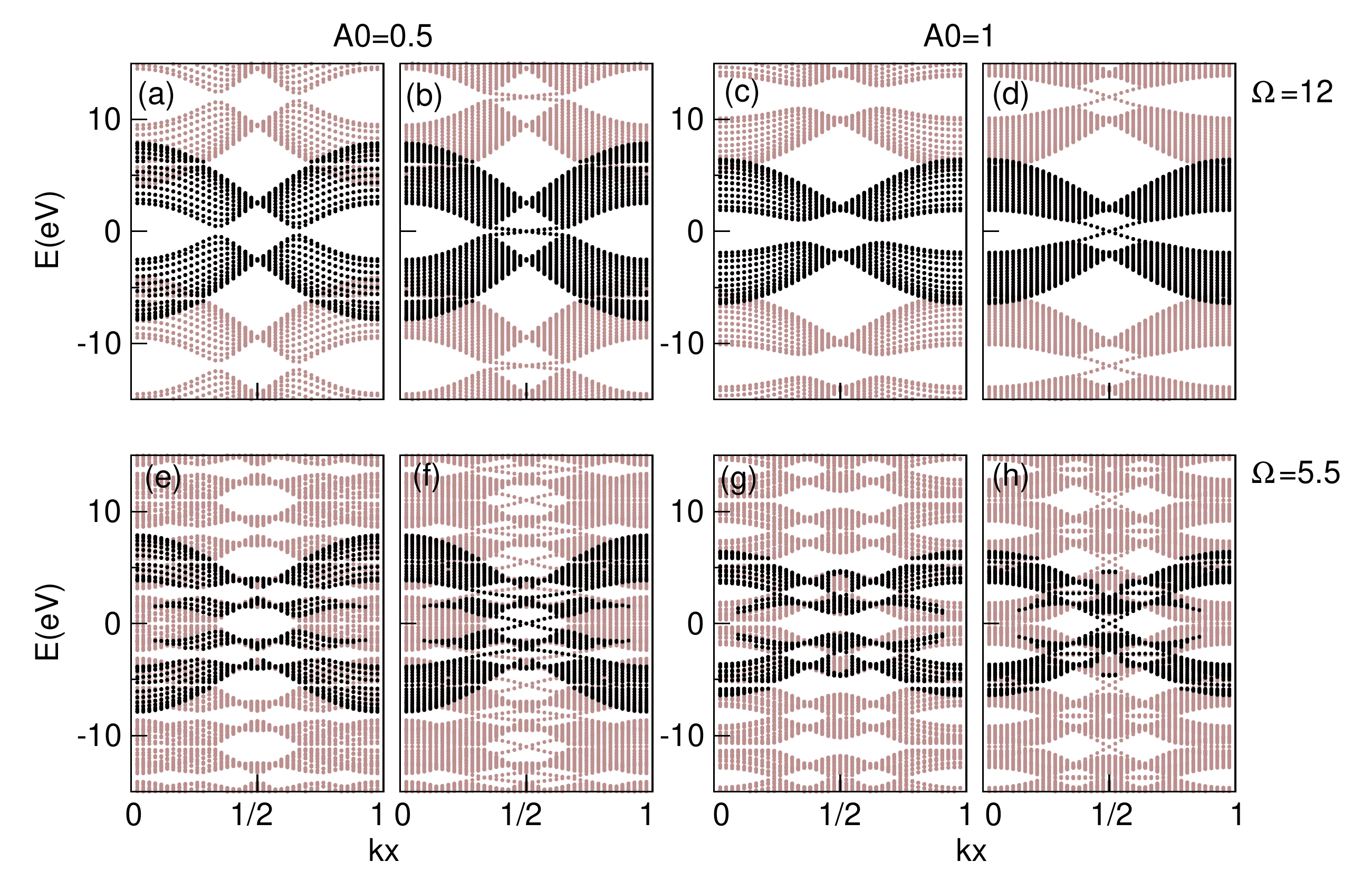}
\caption{ \label{FloqA1}  Floquet quasi-energies  for circularly polarized  field  of different strengths and frequencies.  Panels (a), (c), (e) and (g) report the Floquet Projected Bulk Band Structure, panels (b),(d), (f) and (h)  report the results for the zig-zag ribbon.  The full Floquet band structure (brown dots) corresponds to a replication  of the zero-mode Floquet band (black dots). Upper panels $\Omega=12 eV $, lower panels $\Omega=5.5 eV$,    panels (a),(b),(e) and (f): $A_0=0.5$, panels (c), (d), (g) and (h):  $A_0=1$.}
\end{figure}
\end{center}
\twocolumngrid
We are interested  on the effects of  reduced dimensionality, namely on the gapless edge states that  arise in  graphene ribbons;   we chose a zig-zag terminated ribbon 50-atom wide (Fig. (\ref{geo})).  We consider two frequency values ($\Omega_1= 5.5 eV, \Omega_2=12 eV$) representative of the  intermediate and large frequency regime ($\Omega_1/J \simeq 2 , \Omega_2/J \simeq 4$)  .  We study also the effect of different  amplitudes of the external vector potential ($A_0 = 0.5$ and $A_0 =1.5$ in units of the inverse carbon - carbon distance \cite{Sentef2015}). In  Figs. (\ref{FloqA1}) we compare  the Floquet quasi-energies obtained for the honeycomb lattice  in 2D and  1D  exposed to a circularly polarized field  $\textbf{A}(t)=A_0( cos(\Omega t)\hat{i}+  sin(\Omega t) \hat{j})$.

Panels (a), (c), (e) (g)   report the Floquet Projected   Bulk Band Structure (FPBBS), namely the  Floquet eigenvalues obtained for the 2D lattice  using  the ribbon  unit cell  but adding an extra lattice vector (see Fig. (\ref{geo}) ) to restore the 2D translation symmetry. As currently done in standard surface physics \cite{Manghi1987}, the projected  bulk band structure  allows to identify straightaway  the energy regions that,  prohibited in the bulk,   can host localized   states at the edges. Panels (b), (d), (f), (h) of Fig.  (\ref{FloqA1}) report the Floquet quasi energies obtained in the 1D ribbon geometry clearly showing extra states in the 2D forbidden regions. These states go in pairs being localized either on the upper or on the lower edge of the ribbon. We notice that for  the largest frequency $\Omega= 12 eV$  the effect of increasing $A_0$ is to widen the   gap between bulk Floquet bands and to increase the edge state dispersion. Gapless edge states appearing around $k_{x}=1/2$ (in units of $2 \pi/a$)  exhibit for both values of $A_0$ the peculiar linear dispersion evocative of a non-trivial topological character \cite{PhysRevLett.95.146802,Grandi2015,Grandi2015NJP}.   The same dispersion exists also for  $\Omega= 5.5 eV$   but now the structure of Floquet bands is more complex and extra edge states appear at different k-points and in other gaps and lenses of the FPBBS. Smaller field strengths would correspond to even smaller bulk gaps and less dispersive edge states.

\begin{center}
\begin{figure}[h]
\centering \includegraphics[width=14cm]{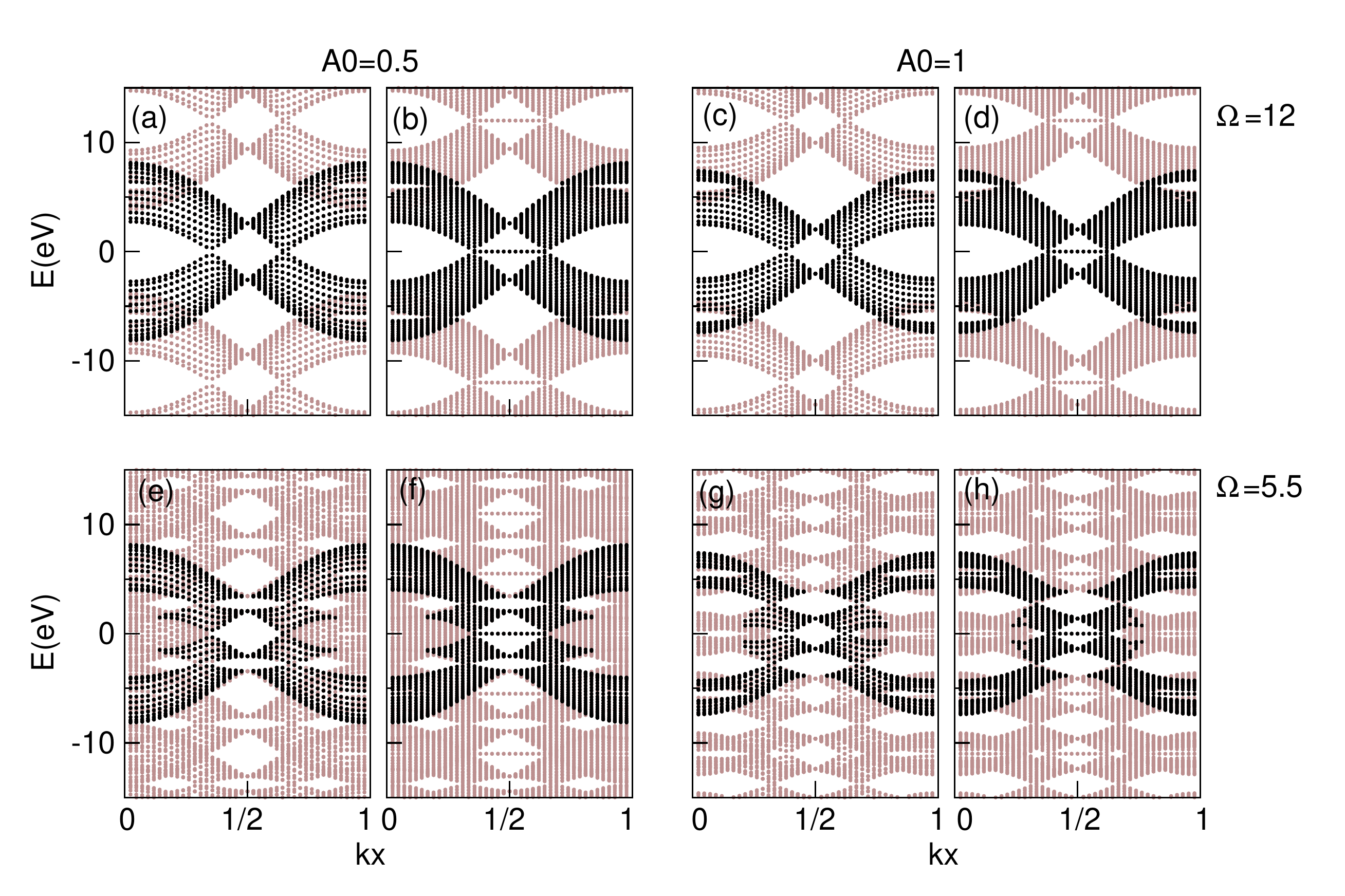}
\caption{\label{FloqA2}  Same as Fig. \ref{FloqA1} but for linear polarization .}
\end{figure}
\end{center}

It is interesting to compare these results with those obtained assuming a linear polarization, with the vector potential oscillating perpendicularly to the ribbon length ( $\textbf{A}(t)=A_0  sin(\Omega t) \hat{j} $).  As shown in Fig. (\ref{FloqA2})  2D Floquet bands are gapless independently on the field strength.  Edge states appearing in the middle of the 1D Brillouine Zone (BZ) have no appreciable k-dispersion  in the same way as  edge states in static conditions \cite{PhysRevLett.102.096801}.
\onecolumngrid
\twocolumngrid
We may conclude this analysis of Floquet quasi-energies by noticing that  only circularly polarized fields of sufficient strength  may induce Floquet edge states with a significant linear dispersion. It is interesting now to go one step further and use the solution of the Floquet problem to obtain information on physical observables.

Even if Floquet quasi-energies  are   used to interpret the  photon dressed electronic excitations  probed by spectroscopic techniques such as  pump-probe experiments and time-resolved photoemission, \cite{Mahmood2016,PhysRevB.94.155304,Sentef2015,DeGiovannini2016} their connection with measurable quantities is somewhat indirect: Floquet quasi-energies  are time-independent eigenvalues of an auxiliary
Hamiltonian and as such cannot be strictly interpreted as "true" observables of the
full time-dependent one. However Floquet eigenvalues and eigenvectors   can be used to calculate expectation values exactly, thanks to the exact representation of the time-dependent
wavefunctions (eq. \ref{FT}). In particular the time-dependent single particle energies of the driven system, defined as the expectation values of the time-dependent
Hamiltonian over $\psi(r,t)$, can be expressed in terms of the Floquet eigenstates as follows
\begin{eqnarray}\label{Edit}
E_{\alpha}(t) &\equiv& <\psi(\textbf{r},t)|\hat{H}(\textbf{r},t)|\psi(\textbf{r},t)> \\ \nonumber
&=& <\psi(\textbf{r},t)|i\frac{\partial}{\partial t}(\textbf{r},t)|\psi(\textbf{r},t)> \\\nonumber
&=& \epsilon_{\alpha}+\sum_{n,m}\sum_{i} e^{i(n-m) \Omega  t} m \Omega <B_{\alpha n}(\textbf{r})|B_{\alpha m}(\textbf{r})>
\end{eqnarray}

These time-dependent single-particle energies represent the extension of the band structure  to the time domain. Notice that at each time step we are able to populate the single particle levels quite naturally by a straightforward electron counting and so to distinguish between filled and empty states   in the same way as in static conditions. On the contrary,  this distinction is far from being obvious for Floquet quasi-energies and the issue of Floquet state occupation remains an open question \cite{PhysRevB.91.235133,PhysRevX.5.041050,Kohn2001,Puviani2015}.

\begin{center}
\begin{figure}[ht]
\centering \includegraphics[width=18cm]{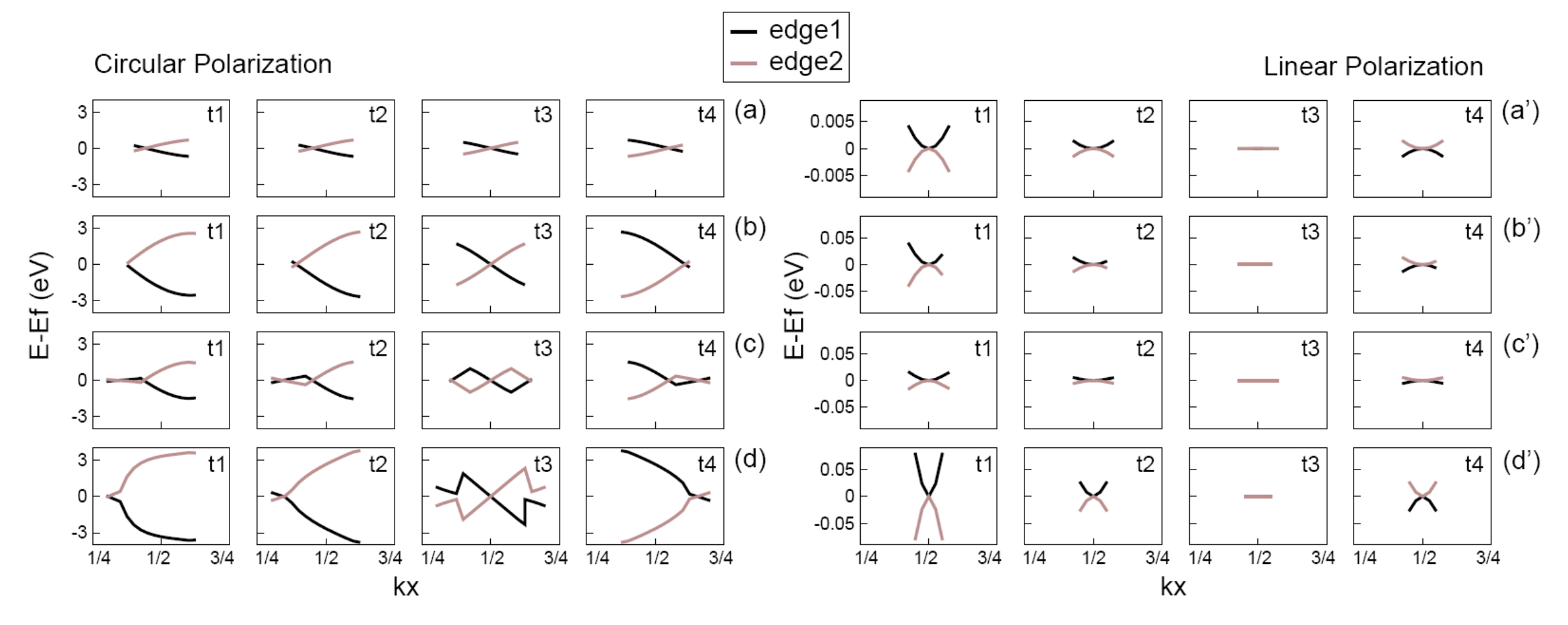}
\caption{\label{edge}  Time evolution of the edge states of a  zig-zag ribbon  for   circular (left panel) and linear polarization (right panel). States localized at upper and lower edge are indicated by black and brown lines respectively. Panels (a), (a'):  $\Omega=12 eV $, $A_0=0.5$; panels (b), (b'): $\Omega=12 eV $, $A_0=1$; panels (c),(c'): $\Omega=5.5 eV $, $A_0=0.5$.   Snapshots    at times
$t_n=(n-1)/8 T$ are reported, $T$ being the  period of the external field.}
\end{figure}
\end{center}

For 1D ribbons we are interested in particular on the time evolution of edge states.    As shown in Figs. (\ref{FloqA1},\ref{FloqA2})  Floquet edge states exist only in a narrow portion of the 1D BZ where we now plot their time evolution (Fig. (\ref{edge})). By looking at the  site composition of the wave function we can unambiguously identify states localized either at the upper or at the lower edge and in the following we  describe the time evolution of each of them.
We consider few  snapshots at selected times $t_n=(n-1)/8 T$ within the interval $0\leq t \leq T/2$, $T $ being the  period of the external field \footnote{Since  $E_{\alpha}(t)= E_{\alpha}(T-t) $ these snapshots are illustrative of the full time evolution.}. We observe that during the time evolution  states localized on a given edge  change their occupation, crossing the Fermi level defined by the above mentioned occupation criterion.
\onecolumngrid
\twocolumngrid
Let us focus first on circular polarization. The high frequency regime (Fig. (\ref{edge}, panels (a) and (b) ) is particularly interesting: during the time evolution the two edge states modify their k-dispersion keeping however the same positive or negative slope:  positive  for states localized at the upper edge, negative for those localized at the lower one. This is remarkable since it corresponds to unidirectional edge states (right movers on the upper edge and left movers on the lower one for  clockwise circular polarization)  that would carry, if occupied, a constant current around the sample \cite{Dahlhaus2015}.  For a smaller frequency (Fig. (\ref{edge}, panels (c), (d)) this is no more completely true and edge states  exhibit a more complex k-dispersion with kinks and variable slopes. We also stress that for smallest frequency and largest intensity (panel (d) of Fig. (\ref{edge})
the edge localization is  less pronounced than in all other cases.

The effect of linearly polarized fields  is significantly different: edge states are confined in a reduced k-space region and  their overall  dispersion is drastically reduced by orders of magnitude with respect to the case of circular polarization. Moreover, and even more notably,  the right mover/left mover behaviour is lost and for any value of frequency and intensity the two edge  bands have a parabolic dispersion,  with  both positive and negative slope. The width of the parabolas varies with time and the  upper/lower edge bands are either fully occupied or empty.

\begin{center}
\begin{figure}[ht]
\centering \includegraphics[width=18cm]{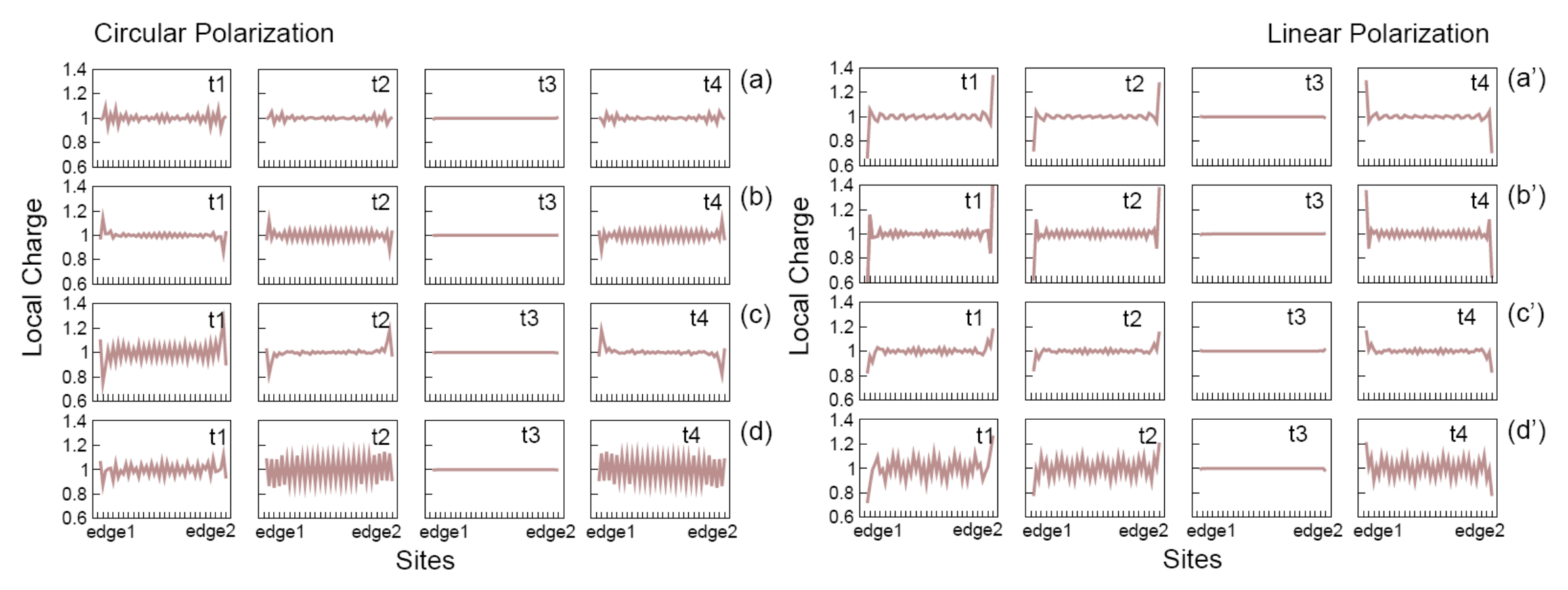}
\caption{\label{rho}   Time evolution of the local charge  calculated for
for circularly  (left panels) and  linearly (right panels) polarized fields of different strengths and frequencies.
Panels (a), (a'): $\Omega= 12 eV$, $A_0=0.5 $.
Panels (b), (b'): $\Omega= 12 eV$, $A_0=1 $.
Panels (c), (c'): $\Omega= 5.5 eV$, $A_0=0.5 $.
Panels (d), (d'): $\Omega= 5.5 eV$, $A_0=1 $.  Snapshots are given at  times $t_n=(n-1)/8 T$.  }
\end{figure}
\end{center}

Since in both conditions of circular and linear polarization the edge state occupation varies in time we expect this to affect the local charge density evolution.  The charge density at each site $i$ in the unit cell is calculated quite simply as the sum of  $|\psi(r,t)|^2$ over the occupied states.
The results for circular and linear polarization are reported in Figs. (\ref{rho}) in units of electrons per site
(assuming 1 valence electron per site in static conditions). Within the ribbon and at the edges charge  oscillates in time as expected; moreover excess charge accumulates at the ribbon edges, moving in time from one edge to the other.

The width of  charge oscillations at the edges is more pronounced for linear polarization and grows  with field intensity, in agreement with physical intuition. The situation is more complex for circular polarization where the strongest oscillations occur for $\Omega= 5.5 eV$, $A_0=0.5$ (panel (c) of Fig. (\ref{rho})). For the same frequency but for higher intensity ((panel (d) of Fig. (\ref{rho})) we notice no  charge  accumulating at the edges. This is a consequence of the poor edge localization occurring for these parameter values.

This analysis allows us to conclude that  fields of appropriate  frequency and intensity of both linear and circular polarization   induce an oscillating dipole across the ribbon. A quantitative analysis of the spectrum of the emitted radiation \cite{Faisal1997} that would involve the current density and its expectation value will be the subject of a further study.
\onecolumngrid
\twocolumngrid
In summary, we have shown how the time evolution of physical observables in systems driven out of equilibrium by a time-periodic electromagnetic field can be obtained from Floquet eigenstates and eigenvalues.
The expectation values of the time-dependent Hamiltonian over the time-dependent single-particle wavefunctions represent an  extension of the band structure to the time domain giving information on the time evolution of single-particle energies, on their population and to  physical quantities that require a summation over occupied states. In graphene ribbons the effects depend strongly on the polarization of the applied field and in the case of circularly polarized light  in a given regime of frequency and intensity,  unidirectional edge states are identified that describe electrons moving in opposite directions along the edges.


%

\end{document}